\documentclass[aps,prb,amsfonts,amssymb,twocolumn,amsmath,preprintnumbers,nofootinbib,floatfix,showpacs]{revtex4}
\usepackage{amsmath,bm,amsfonts,amssymb}
\usepackage[dvips]{graphics}
\usepackage{graphicx}

\begin{document}

\title{Subharmonic Gap Structure in Superconductor/Ferromagnet/Superconductor Junctions}
\author{I. V. Bobkova}
\email[Electronic address: ]{bobkova@issp.ac.ru}
\affiliation{Institute of Solid State Physics, Chernogolovka,
Moscow reg., 142432 Russia}

\date{\today}

\begin{abstract}
The behavior of dc subgap current in magnetic quantum point
contact is discussed for the case of low-transparency junction
with different tunnel probabilities for spin-up ($D_\uparrow$) and
spin-down ($D_\downarrow$) electrons. Due to the presence of
Andreev bound states $\pm \varepsilon_0$ in the system the
positions of subgap electric current steps $eV_n = (\Delta \pm
\varepsilon_0)/n$ are split at temperature $T \neq 0$ with respect
to the nonmagnetic result $eV_n=2\Delta/n$. It is found that under
the condition $D_\uparrow \neq D_\downarrow$ the spin current also
manifests subgap structure, but only for odd values of $n$. The
split steps corresponding to $n=1,2$ in subgap electric and spin
currents are analytically calculated and the following steps are
described qualitatively.
\end{abstract}

\maketitle

It is well established now that subharmonic gap structure in
superconducting weak links can be described in terms of multiple
Andreev reflection (MAR). The concept of MAR was first introduced
\cite{Klapwijk82} for SNS (superconductor - normal metal -
superconductor) junctions and then extended to include the effect
of resistance of SN interface \cite{Octavio8388} (OTBK theory).
Then it was shown \cite{Arnold} that this mechanism is the reason
for subharmonic gap structures in junctions of any type, and
specifically in superconducting point contacts. As long as the
junction is short on the scale of the coherence length $\xi_0$ the
microscopic details of the junction are irrelevant and for the
nonmagnetic junction the only parameter is the transmission
probability of the barrier $D$. An exact theory of MAR for short
nonmagnetic junctions between clean superconductors has been
developed in Refs.\onlinecite{Shumeiko95, Averin95} on the basis
of scattering amplitudes method and in Ref.\onlinecite{Cuevas96}
making use of nonequilibrium Green function techniques. It was
obtained that for this case MAR manifests itself in dc current as
current steps at voltages $eV_n=2\Delta/n$, $n=1,2,...$. The
magnitude of the current onset at $V_n$ is proportional to $D^n$.
The theoretical results are consistent with the experimental
situation in atomic-size superconducting point contacts
\cite{vanderPost94, Scheer97, Scheer98, Ludoph00}. The subharmonic
gap structure due to photon-assisted MAR in the presence of
microwave radiation has also been studied and theoretical
predictions \cite{Cuevas98} are confirmed by the experimental
observations \cite{Uzawa05}.

Arnold \cite{Arnold} was the first to take into account the phase
coherence of subsequent Andreev reflections, which were absent in
the OTBK theory. The quantum coherence in short ballistic
junctions lead to ac Josephson effect \cite{Shumeiko95, Averin95,
Cuevas96}. In addition, in long ballistic SNS junctions coherence
effects give rise to resonant structures in the dc current due to
Andreev quantization \cite{Shumeiko01}.

MAR in diffusive nonmagnetic junctions has also been investigated
in details. Short diffusive junctions with Thouless energy $E_{Th}
\gg \Delta$ can be described by coherent MAR theory and show
subharmonic gap structures at $eV_n=2\Delta/n$ \cite{Bardas97,
Zaitsev98}. In long diffusive junctions with $E_{Th} \lesssim
\Delta$ additional peaks in the differential conductance at $eV_n
\approx 2(\Delta \pm E_{Th})/2n$ are predicted \cite{Shumeiko02}.
This result is in qualitative agreement with the experimental data
\cite{Kutchinsky97}. The incoherent MAR regime with $eV, \Delta
\gg E_{Th}$ has also been studied \cite{Bezuglyi99, Bezuglyi00}
and it was found that subharmonic gap structure shows
qualitatively different behavior for even and odd subharmonic gap
structures.

Magnetic interfaces can not be characterized by the only parameter
$D$ and Andreev bound states are known to take place on the
magnetic surface of s-wave superconductor \cite{Fogelstrom00}. So
it is natural that new qualitative features with respect to
nonmagnetic case appear in dc current even in the simplest models
of a short magnetic junction \cite{Cuevas01,Fogelstrom02}. In
Ref.\onlinecite{Fogelstrom02} the dc-current-voltage
characteristic of magnetic quantum point contact between two
superconductors is studied theoretically for the model where
magnetic quantum dot is characterized by two parameters:
transparency $D$ and the spin-mixing angle $\Theta$. It is found
that subgap current steps take place at voltages $eV_n =
\Delta(1+\cos \Theta/2)/n$.

In this paper the dc subgap current due to MAR processes in short
ballistic quantum point contact for the most general case of
symmetric magnetic interface including different transparencies
for spin-up and spin-down quasiparticles ($D_\uparrow \neq
D_\downarrow$) is studied theoretically. I consider the case of
small $D_\uparrow$ and $D_\downarrow$ of the same order of
magnitude and find that at $T \neq 0$ current steps occur at
voltages $eV_n = \Delta(1 \pm \cos \Theta/2)/n$. This splitting
disappears in the limit $T \to 0$ and the result obtained in
Ref.\onlinecite{Fogelstrom02} for the current step positions is
recovered. Also I found that the difference between $D_\uparrow$
and $D_\downarrow$ is very important. It leads to non-zero spin
current which has subharmonic gap structure different from that
one for the dc electric current.

The theoretical analysis is based on the non-equilibrium
quasiclassical theory of superconductivity in terms of Riccati
amplitudes developed by Eschrig \cite{Eschrig00} and generalized
for the case of magnetic interfaces in
Refs.\onlinecite{Fogelstrom00, Zhao04}. Now I briefly outline this
formalism. The fundamental quantity in non-equilibrium
quasiclassical theory of superconductivity is the quasiclassical
Green's function $\check g = \check g(\bm p_f, \bm R, \epsilon,
t)$. It is a $8\times8$ matrix form in the product space of
Keldysh, particle-hole and spin variables. In general the
quasiclassical Green's functions are depend on space $\bm R$, time
$t$ variables, the direction of quasiparticle Fermi momentum $\bm
p_f$ and the excitation energy $\epsilon$. In our case of one-mode
quantum point contact the problem is effectively one-dimensional
and $\bm R \equiv x$, where $x$ - is the coordinate measured along
the normal to the junction. The momentum $\bm p_f$ has only two
values, which correspond to incoming and outgoing trajectories.

The electric and spin currents should be calculated via Keldysh
part of the quasiclassical Green's function. For the one-mode
quantum point contact the corresponding expression for the charge
current reads as follows
\begin{eqnarray}
&\frac{\displaystyle j^{ch}}{\displaystyle e} =
\frac{\displaystyle {\rm sgn} \bm p_f}{\displaystyle 2 \pi}
\displaystyle \int \limits_{-\infty}^{+\infty} \frac{\displaystyle
d \epsilon}{\displaystyle 4 \pi i } \times \qquad \qquad \qquad
\qquad \qquad
\nonumber \\
&{\rm Tr}_4 \left[\hat \tau_3 \hat \sigma_{0} \left(\check g^K(\bm
p_f, x, \epsilon, t)-\check g^K(\underline{\bm p}_f, x, \epsilon,
t)\right)\right] \label{current} \enspace ,
\end{eqnarray}
where $e$ is the electron charge and $\hbar = 1$ throughout the
paper. $\check g^K(\bm p_f, x, \epsilon, t)$ is a $4\times4$
Keldysh Green's function in the product space of particle-hole and
spin variables. $\bm p_f$ stands for incoming quasiparticle
trajectories and $\underline{\bm p}_f$ for the outgoing ones.
$\hat \tau_i$ and $\hat \sigma_i$ are Pauli matrices in
particle-hole and spin spaces, respectively. The spin current
$j^{sp}/s^e$ can be calculated making use of Eq.(\ref{current})
with the substitution $\hat \sigma_3$ for $\hat \sigma_0$.
$s^e=1/2$ is the electron spin.

Quasiclassical Green's function $\check g$ can be expressed in
terms of Riccati coherence functions $\hat \gamma^{R,A}$ and $\hat
{\tilde \gamma}^{R,A}$, which measure the relative amplitudes for
normal-state quasiparticle and quasihole excitations, and
distribution functions $\hat x^K$ and $\hat {\tilde x}^K$. All
these functions are $2 \times 2$ matrices in spin space and depend
on $(\bm p_f, x, \epsilon, t)$.

In this paper I consider a short junction with the characteristic
size $d \ll \xi_0$. But it is assumed that despite the small size
of the grain charging effects can be neglected. For definiteness
the current is calculated on the left side of the interface.
Keldysh Green's function for incoming trajectory is parameterized
by
\begin{widetext}
\begin{equation}
\check g_1^K(\bm p_f) = -2 i \pi \check N^R \otimes
\left(
\begin{array}{cc}
(\hat x_1^K -\hat \gamma_1^R  \otimes \hat {\tilde X}_1^K \otimes
\hat {\tilde \gamma}_1^A ) & -(\hat \gamma_1^R  \otimes \hat
{\tilde
X}_1^K -\hat x_1^K  \otimes \hat \Gamma_1^A)  \\
-(\hat {\tilde \Gamma}_1^R  \otimes \hat x_1^K -\hat {\tilde
X}_1^K \otimes \hat {\tilde \gamma}_1^A) & (\hat {\tilde X}_1^K
-\hat {\tilde \Gamma}_1^R \otimes x_1^K \otimes \hat \Gamma_1^A )
\end{array}
\right) \otimes \check N^A \label{g_Riccati} \enspace ,
\end{equation}
\begin{equation}
\check N^{R(A)} = \left(
\begin{array}{cc}
\left( 1 - \hat \gamma_1^R (\hat \Gamma_1^A) \otimes \hat {\tilde
\Gamma}_1^R (\hat {\tilde \gamma}_1^A) \right)^{-1} & 0 \\
0 & \left( 1 - \hat {\tilde \Gamma}_1^R (\hat {\tilde \gamma}_1^A)
\otimes \hat \gamma_1^R (\hat \Gamma_1^A) \right)^{-1}
\end{array}
\right) \enspace . \label{Ng}
\end{equation}
\end{widetext}
Here subscript 1 means that the corresponding functions should be
taken at $x=-0$, argument $\bm p_f$ of all the Riccati functions
is omitted for brevity. The product $\otimes$ of two functions of
energy and time is defined by the noncommutative convolution $A
\otimes B =
e^{i(\partial_\epsilon^A\partial_t^B-\partial_t^A\partial_\epsilon^B)}A(\epsilon,t)B(\epsilon,t)$.
Keldysh Green's function $\check g_1(\underline{\bm p}_f)$ for the
outgoing trajectory can be obtained from Eqs. (\ref{g_Riccati}),
(\ref{Ng}) by the substitution $(\hat \gamma_1^R, \hat {\tilde
\gamma}_1^A, \hat x_1^K)(\bm p_f) \to (\hat \Gamma_1^R, \hat
{\tilde \Gamma}_1^A, \hat X_1^K)(\underline{\bm p}_f)$ and $(\hat
{\tilde \Gamma}_1^R, \hat \Gamma_1^A, \hat {\tilde X}_1^K)(\bm
p_f) \to (\hat {\tilde \gamma}_1^R, \hat \gamma_1^A, \hat {\tilde
x}_1^K)(\underline{\bm p}_f)$.

Riccati coherence and distribution functions obey Riccati-type
transport equations. The equations are given in
Refs.\onlinecite{Eschrig00,Zhao04}, so I do not write them here.
The quantities $(\hat \gamma_1^{R,A}, \hat {\tilde
\gamma}_1^{R,A}, \hat x_1^K, \hat {\tilde x}_1^K)$, denoted by
lower case symbols, are obtained by solving the Riccati equations
for the appropriate trajectory with the asymptotic conditions,
which are for spin-singlet s-wave superconductor as follows
\begin{equation}
\hat \gamma_{l,r}^{R,A}(\epsilon,t)= \left\{
\begin{array}{ll}
\frac{\displaystyle \Delta e^{-2 i e V_{l,r}t}}{\displaystyle
\epsilon \pm i
\sqrt{\Delta^2-\epsilon^2}}i \hat \sigma_2, & |\epsilon|<\Delta \\
\frac{\displaystyle \Delta e^{-2 i e V_{l,r}t}}{\displaystyle
\epsilon + {\rm sgn \epsilon}
\sqrt{\epsilon^2-\Delta^2}}i \hat \sigma_2, & |\epsilon|>\Delta \enspace , \\
\end{array}
\right. \label{gamma_asympt}
\end{equation}
\begin{equation}
\hat x_{l,r}^{K}(\epsilon) = \left( 1-|\hat
\gamma_{l,r}^R((\epsilon-eV_{l,r}),t)|^2 \right)\tanh
\frac{\epsilon -e V_{l,r}}{2 T} \label{x_asympt} \enspace ,
\end{equation}
where the subscript $l,r$ denotes that the appropriate Riccati
function belongs to the bulk of the left (right) superconductor.
$\Delta$ is the bulk absolute value of superconducting order
parameter for a given temperature, which is assumed to be the same
in the both superconductors. $V_{l,r}$ is the electric potential
in the bulk of left (right) superconductor, so $V=V_r-V_l$ is the
voltage bias applied to the junction. Quantities $\hat {\tilde
\gamma}_{l,r}^{R,A}$ and $\hat {\tilde x}_{l,r}^K$ are obtained
from Eqs. (\ref{gamma_asympt}), (\ref{x_asympt}) respectively by
the operation $\tilde a(\epsilon, t) = a(-\epsilon, t)^*$.

For analytical consideration superconducting order parameter and
electric potential are assumed to be spatially constant in the
superconductors. Under this assumption the voltage drop only
occurs at the junction region. These simplifications seem to be
reasonable for quantum point contact. Under the assumptions above
the solutions of Riccati equations for $(\hat \gamma^{R,A}(x),
\hat {\tilde \gamma}^{R,A}(x), \hat x^K(x), \hat {\tilde x}^K(x))$
do not depend on the space variable and coincide with the
asymptotic conditions in the corresponding superconductor.

The quantities $(\hat \Gamma_1^{R,A}, \hat {\tilde
\Gamma}_1^{R,A}, \hat X_1^K, \hat {\tilde X}_1^K)$, denoted by
upper case symbols, are expressed via $(\hat \gamma_1^{R,A}, \hat
{\tilde \gamma}_1^{R,A}, \hat x_1^K, \hat {\tilde x}_1^K) = (\hat
\gamma_l^{R,A}, \hat {\tilde \gamma}_l^{R,A}, \hat x_l^K, \hat
{\tilde x}_l^K)$ and the elements of the interface scattering
matrix $\cal S$ for the normal-state electrons and holes with the
energies at the Fermi surface. The interface $\cal S$-matrix is a
unitary $8 \times 8$ matrix in the combined spin, particle-hole
and directional spaces. The explicit structure of $\cal S$-matrix
in directional space is
\begin{equation}
\cal S =
\left(
\begin{array}{cc}
\check S_{11} & \check S_{12} \\
\check S_{21} & \check S_{22} \\
\end{array}
\right)
\label{S_general} \enspace ,
\end{equation}
where matrix $\check {S}_{ii}$ contains spin-dependent reflection
amplitudes of normal-state quasiparticles from the interface in
$i$-th half-space, while $\check {S}_{ij}$ with $i\ne j$
incorporates spin-dependent transmission amplitudes of
normal-state quasiparticles from side $j$. Each element $\check
S_{ij}$ is a diagonal matrix in particle-hole space $\check S_{ij}
= \hat S_{ij} (1 + \hat \tau_3)/2+\hat {\tilde S}_{ij}(1 - \hat
\tau_3)/2$. The most general form of {\cal S}-matrix for a
symmetric magnetic interface without spin-orbit interaction can be
written as \cite{bb02}:
\begin{equation}
\hat S_{11} = \hat S_{22} =
\left(
\begin{array}{cc}
\sqrt {R_\uparrow}e^{i \Theta/2} & 0 \\
0 & \sqrt {R_\downarrow}e^{-i \Theta/2} \\
\end{array}
\right) \label {S_ii} \enspace ,
\end{equation}
\begin{equation}
\hat S_{12} = \hat S_{21} = \pm i
\left(
\begin{array}{cc}
\sqrt {D_\uparrow}e^{i \Theta/2} & 0 \\
0 & -\alpha \sqrt {D_\downarrow}e^{-i \Theta/2} \\
\end{array}
\right) \label {S_ij} \enspace ,
\end{equation}
where $R_{\uparrow,\downarrow} + D_{\uparrow,\downarrow} = 1$,
$\alpha = \pm 1$ depending on the particular model \cite{bb02}.
$\hat {\tilde S}_{ij} = \hat S_{ij}$ in the considered problem.
The particular expressions for $(\hat \Gamma_1^{R,A}, \hat {\tilde
\Gamma}_1^{R,A}, \hat X_1^K, \hat {\tilde X}_1^K)$ in terms of
$(\hat \gamma_1^{R,A}, \hat {\tilde \gamma}_1^{R,A}, \hat x_1^K,
\hat {\tilde x}_1^K)$ and $\cal S$-matrix elements can be found in
Ref.\onlinecite{Zhao04}, so I do not write them explicitly.

\begin{figure*}[!tbh]
\begin{minipage}[b]{.5\linewidth}
   \centerline{\includegraphics[clip=true,width=3in]{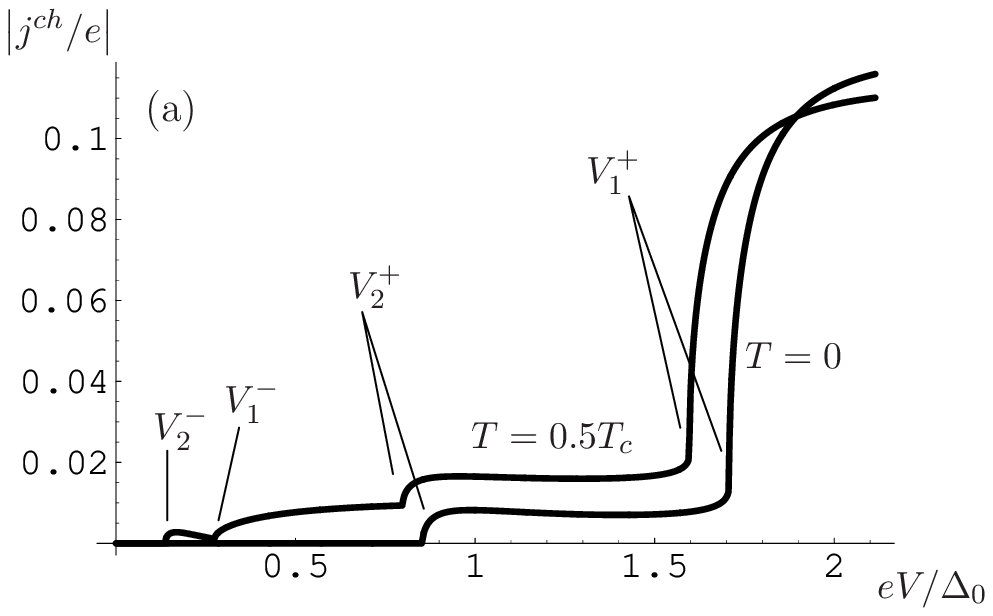}}
  \end{minipage}\hfill
  \begin{minipage}[b]{.5\linewidth}
   \centerline{\includegraphics[clip=true,width=3in]{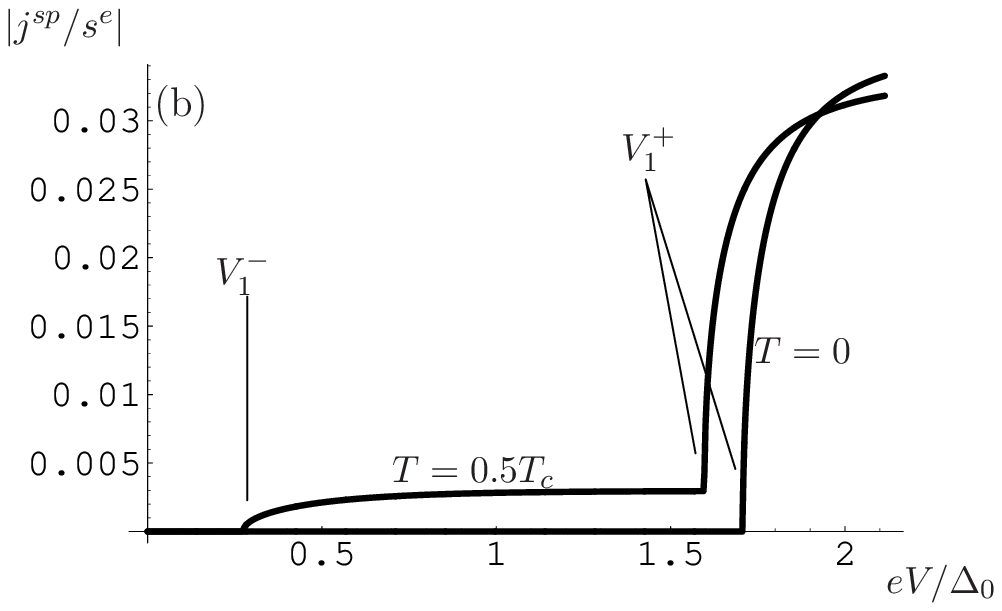}}
  \end{minipage}\hfill
\caption{(a) The subgap structure of dc electric current (in units
of $\Delta_0 e/\hbar$) is plotted in accordance with
Eq.({\ref{j_general_tun}}) for $T=0,0.5T_c$. (b) The same for the
spin current. $\Theta = \pi/2$, $D_\uparrow = 0.2$, $D_\downarrow
= 0.1$. $\Delta_0 \equiv \Delta(T=0)$.} \label{fig1}
\end{figure*}

Now we can proceed with the current. Substituting Keldysh Green's
function (\ref{g_Riccati}) into Eq.(\ref{current}) it can be seen
that the current is expressed as a sum over harmonics $j(t) = \sum
\limits_N J_N e^{2 i N e V t}$. I focus on the dc component and
only calculate the term corresponding to $N=0$. I-V
characteristics in charge and spin currents exhibit steps. In the
tunnel limit one can extract the elementary processes giving rise
to each of these steps. In this paper the positions, height and
shape of several first steps in subgap I-V characteristic are
calculated analytically. The linear in transparency
thermally-activated term in the dc current, which is not related
to the steps, has also been calculated, but it is not pronounced
in the subgap region for the considered problem and is not
discussed below. The part of the dc current corresponding to the
steps takes the form:
\begin{equation}
\frac{j^{ch}}{e}(\frac{j^{sp}}{s^e}) =
j_{1+}^{ch(sp)}+j_{1-}^{ch(sp)}+j_{2+}^{ch(sp)}+j_{2-}^{ch(sp)}
\label{j_general_tun} \enspace ,
\end{equation}
\begin{widetext}
\begin{equation}
j_{1+}^{ch(sp)} = \frac{\Delta |\sin \Theta/2|}{4}(D_\uparrow \pm
D_\downarrow) \frac{|\varepsilon_0 - e V |\sqrt{(\varepsilon_0 - e
V)^2-\Delta^2}}{(\varepsilon_0 - e V)^2-\Delta^2 \cos^2 \Theta/2}
\left( \tanh \frac{\varepsilon_0 - e V}{2 T} - \tanh
\frac{\varepsilon_0}{2 T} \right) \label{j_1_tun} \enspace ,
\end{equation}
\begin{equation}
j_{2+}^{ch}= \frac{\Delta |\sin \Theta/2|}{2}D_\uparrow
D_\downarrow \frac{|\varepsilon_0 - 2 e V |\sqrt{(\varepsilon_0 -
2 e V)^2-\Delta^2}}{(\varepsilon_0 - 2 e V)^2-\Delta^2 \cos^2
\Theta/2} \frac{1 + |\gamma(\varepsilon_0 -  e V )|^2}{|1 -
\gamma^2(\varepsilon_0 - e V)e^{i \Theta}|^2} \left( \tanh
\frac{\varepsilon_0 - 2 e V}{2 T} - \tanh \frac{\varepsilon_0}{2
T} \right)\label{current_steps_tun} \enspace ,
\end{equation}
\end{widetext}
$j_{1-}^{ch(sp)} = -j_{1+}^{ch(sp)} (V \to -V)$, $j_{2-}^{ch} =
-j_{2+}^{ch} (V \to -V)$, $j_{2\pm}^{sp} = 0$. In
Eqs.(\ref{current_steps_tun}),(\ref{j_1_tun}) $\varepsilon_0 =
\Delta \cos \Theta/2$.
$\gamma(\epsilon)=(\epsilon-i\sqrt{\Delta^2-\epsilon^2})/\Delta$,
where for $|\epsilon|>\Delta$ the square root
$\sqrt{\Delta^2-\epsilon^2}$ should be substituted by $- i {\rm
sgn} \epsilon \sqrt{\epsilon^2 - \Delta^2}$. The behavior of
$j^{ch}$ and $j^{sp}$ described by Eq. (\ref{j_general_tun}) is
shown in Fig.\ref{fig1} for $\Theta=\pi/2$.

It is seen from Fig.\ref{fig1} that at zero temperature there is a
shift of well-known step positions in electric current for a
non-magnetic interlayer: $eV_n=2\Delta/n \to
eV_n^+=(\Delta+\varepsilon_0)/n$. This result is coincide with
that one found in Ref.\onlinecite{Fogelstrom02}. However, at
non-zero temperature non-magnetic step positions are split:
$eV_n=2\Delta/n \to eV_n^{\pm}=(\Delta \pm \varepsilon_0)/n$. It
is easily obtained that the additional steps $eV_n^-$ disappear
rapidly ($\propto e^{-\varepsilon_0/T}$) when the temperature goes
down. In the first order of $D$ the analytical expression for the
electric current similar to Eq. (\ref{j_1_tun}) has been obtained
for the case of nonmagnetic junction between two d-wave
superconductors \cite{barash97}.

Let us describe all the steps for $n > 2$ qualitatively. The
estimate for the magnitude of $j_n^{ch(sp)}$ is easily obtained if
one considers physical processes of MAR leading to subgap current
steps. For the spin-up band the bound state energy is
$\varepsilon_B=+\varepsilon_0$ and at $T=0$ the following process
gives spin-up current at $eV_n^{+}$: a spin-up electron tunnels
into the bound state from the continuum states below $- \Delta$
undergoing $n-1$ Andreev reflections inside the gap. An electron
converts into a hole and vice versa under each Andreev reflection
event. In every passing of the electron (hole) through the
boundary the amplitude of its wave function should be multiplied
by $\sqrt{D_\uparrow}(\sqrt{D_\downarrow})$. For the spin-down
band with $\varepsilon_B = - \varepsilon_0$ the same process takes
place with the modification that a spin-down electron starts in
the bound state and goes up into the continuum states at $+
\Delta$. So the current $j_{\uparrow, \downarrow 2k}$ is $\propto
(D_\uparrow D_\downarrow)^k$ and $j_{\uparrow, \downarrow (2k+1)}
\propto (D_\uparrow D_\downarrow)^k D_{\uparrow, \downarrow}$.
Finally, we obtain for $j_n^{ch(sp)} = j_{n\uparrow} \pm
j_{n\downarrow}$ the following result: $j_{(2k+1)}^{ch(sp)}
\propto (D_\uparrow D_\downarrow)^k (D_\uparrow \pm D_\downarrow)$
and $j_{(2k)}^{ch} \propto 2(D_\uparrow D_\downarrow)^k$,
$j_{(2k)}^{sp} = 0$.

For $T \neq 0$ there is a non-zero probability to find a spin-up
electron at $\varepsilon_B = \varepsilon_0$ and a vacant place for
a spin-down electron at $- \varepsilon_0$. So the processes
leading to the additional steps at $eV_n^-$ arise: (i) a spin-up
electron tunneling from $\varepsilon_B = \varepsilon_0$ to the
continuum spectrum at $+ \Delta$ and (ii) a spin-down electron
tunneling from the continuum spectrum below $- \Delta$ into the
bound state $-\varepsilon_0$. Consequently, for $T \neq 0$ the
splitting of current step positions occurs and there are all the
steps at $eV_n^{\pm}=(\Delta \pm \varepsilon_0)/n$ in the charge
current and only odd split steps in the spin current.

Of course, the processes of inelastic scattering lead to
impossibility of observing the subgap structures corresponding to
large $n$ due to the loss of coherence if the time
$\tau=nd_{eff}/v_f$, which a quasiparticle spends in the junction
region, exceeds the average time of inelastic scattering
$\tau_{in}$. So the steps with $n$ more than
$\tau_{in}\Delta_0(\xi_0/d_{eff})$ can not be observed. For the
case under consideration this estimate seems not to be restrictive
because $\xi_0/d_{eff} \sim 1$ for a short junction and $\tau_{in}
\Delta_0$ due to electron-phonon scattering $\propto
(\omega_D/T)^2(\Delta_0/T)$ and considerably exceeds unity for
temperatures up to $T_c$.

In conclusion, I have presented a theoretical study of the dc
electric and spin current subgap structure in magnetic point
contact. At $T \neq 0$ electric current step positions
$eV_n=(\varepsilon_0\pm\Delta)/n$ are split compared to the case
of nonmagnetic junction and at $T=0$ they are just shifted:
$eV_n=(\varepsilon_0+\Delta)/n$. The subgap  spin current, taking
place at $D_\uparrow \neq D_\downarrow$ manifests steps at the
same voltages but only for odd values of $n$. The current for
steps with $n=1,2$ is calculated analytically and a qualitative
description of the following steps is presented.

{\it Acknowledgments.} The author thanks A.M. Bobkov for
enlightening discussions and acknowledges the support by grant
RFBR 05-02-17175.

\end{document}